\documentclass{mn2e}
\usepackage{psfig}
\usepackage{epsfig}

\newif\ifAMStwofonts

\def\sqiglt{\hbox{\rlap{\lower.55ex \hbox {$\sim$}}\kern-.05em \raise.4ex \hbox{$<$}\,}}
\def\sqiggt{\hbox{\rlap{\lower.55ex \hbox {$\sim$}}\kern-.05em \raise.4ex \hbox{$>$}\,}}

\def\til{\ensuremath{\sim\,}}

\def\chisq{\ensuremath{\chi^2}}
\def\rchisq{\ensuremath{\chi_{\nu}^{2}}}

\newcommand{\tim}[1]{\ensuremath{\times 10^{#1}}}
\def\deg{\ensuremath{^{\circ}}}
\def\etal{et al.\ }
\def\mekal{{\sc mekal}}
\def\xmmn{\emph{XMM-Newton}}
\def\xmm{\emph{XMM}}
\def\cms{cm$^{-2}$}
\def\cps{counts s$^{-1}$}

\def\rwd{\ensuremath{R_{\rm WD}}}
\def\d{\ensuremath{\delta}}

\title[HT~Cam]{HT~Camelopardalis: The simplest intermediate polar
spin pulse}
\author[Evans \&\ Hellier]{P.A. Evans\thanks{pae@astro.keele.ac.uk} and Coel
Hellier\\ Astrophysics Group, School of Chemistry and Physics, Keele
University, Staffordshire, ST5 5BG}

\date{Accepted 
      Received }

\pagerange{\pageref{firstpage}--\pageref{lastpage}}
\pubyear{2005}

\begin{document}

\maketitle

\label{firstpage}

\begin{abstract}
The intermediate polar HT~Cam is unusual in that it shows no evidence for dense
absorption in its spectrum. We analyse an \xmmn\/\ observation of this star,
which confirms the absence of absorption and shows that the X-ray spin-pulse is
energy independent. The modulation arises solely from occultation effects and
can be reproduced by a simple geometrical model in which the lower accretion
footprint is fainter than the upper one. 

We suggest that the lack of opacity in the accretion columns of HT~Cam, and
also of EX~Hya and V1025~Cen, results from a low accretion rate owing to their
being below the cataclysmic variable period gap.
\end{abstract}



\section{Introduction}
\label{sec:intro}

\begin{figure*}
\begin{center}
\psfig{file=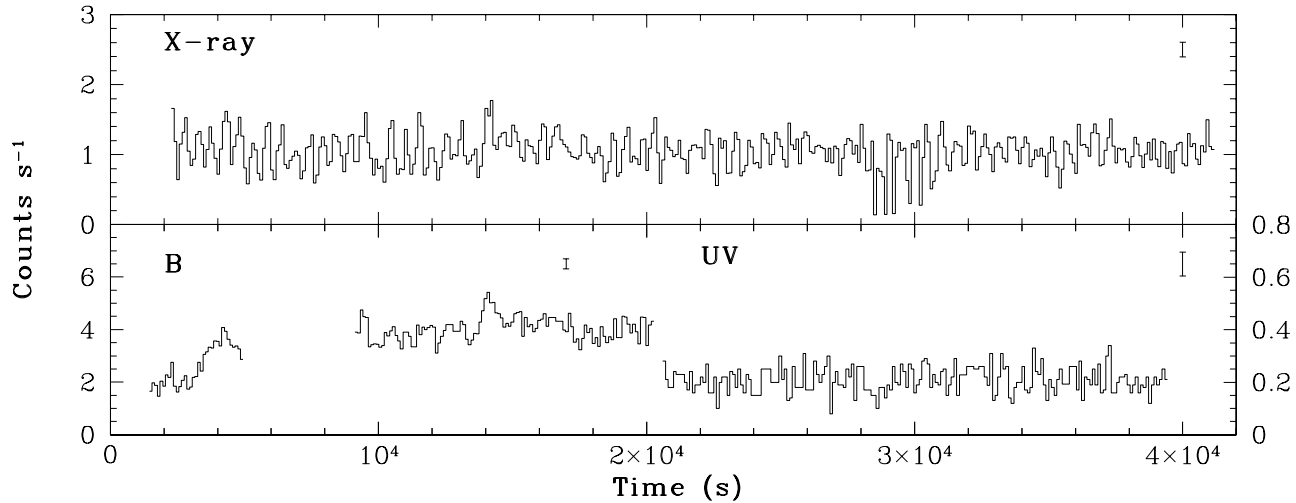,height=17cm,angle=-90}
\caption{Lightcurves of HT~Cam in the X-ray (EPIC-pn; 0.2--12 keV), $B$
(3900--4900 \AA) and UV (1800--2250 \AA) bands. The bins are 100 s, and the
515-s spin-pulse can be clearly seen in the X-ray band. Typical errors are
shown.}
\label{fig:curve}
\end{center}
\end{figure*}

A characteristic feature of intermediate polars (IPs) -- interacting
white-dwarf/red-dwarf binaries with a magnetic primary -- is that their X-ray
emission shows a clear modulation at the white-dwarf spin period (see Patterson
1994 for an overview of these objects). 

Understanding these pulsations, however, is not straight forward since many
factors can contribute to the modulation. These include opacity in the X-ray
emitting accretion regions, and opacity owing to infalling material if it
passes through the line of sight. Further, they include occultation as the
accretion regions pass over the white-dwarf limb, including the effect of
possible asymmetries between the two accreting poles. Thus IP spin pulses can
show complex profiles, as exemplified by FO~Aqr (e.g.\ Beardmore \etal1998) and
PQ~Gem (e.g.\ Mason 1997).

It is thus useful to study systems where we can understand the pulse profile in
terms of simple geometry alone. One such IP is XY~Ari, where an eclipse
provides geometrical information (Hellier 1997). Recent work on HT~Cam suggests
that it has an unusually simple spin pulse, since there appears to be very
little opacity affecting the X-ray emission (de~Martino \etal2005), and thus no
complications owing to absorption by accretion curtains.

In this paper we analyse \xmm\/\ data on HT~Cam and produce a geometric model
of the pulse, leading to constraints on the accretion geometry in this system.


\section{Observations}
\label{sec:obs}

HT~Camelopardalis (RX\,J0757.0+6306, hereafter HT~Cam) was observed by the
\xmmn\/\ satellite (Jansen \etal2001) for \til40 ks on 2003 March 24. AAVSO
data reveal that the system was quiescent at $V\til17$. The EPIC MOS (Turner
\etal2001) and PN (Str\"uder \etal2001) instruments were operating in {\sc full
frame mode}, observing through the Medium filter. Data from these three
instruments were extracted from circular regions of radius 14 arc sec, centred
on the source. An annulus around this was used to estimate the background. The
OM camera (Mason \etal2001) was also in operation, collecting 17.5 ks of data
through each of the $B$ and UVW2 filters. The data were analysed using the {\sc
xmm-sas} software, v6.0.0. 

The lightcurves are shown in Fig.~\ref{fig:curve}. The X-ray data show clear
modulation at the 515-s white-dwarf spin period. Fourier analysis confirms the
spin pulse but shows no orbital or beat modulation. In the $B$ band there is a
\til5 per cent spin-period modulation, but there is no detectable pulse in the
UV-band, up to a limit of \til14 per cent. The mean X-ray count-rate in the pn
detector is 1.05 \cps, which corresponds to a 0.2--12 keV flux of 6.5\tim{-12}
erg \cms\ s$^{-1}$; for comparison the 0.1--2.4 keV flux observed by
\emph{ROSAT\/} was 1.87\tim{-12} erg \cms\ s$^{-1}$ (Tovmassian \etal1998).


\section{Spectroscopy}
\label{sec:spec}

X-ray emission in an IP comes from multi-temperature plasmas beneath the
accretion shock. To model the spectrum of HT~Cam we used \mekal\ components in
{\sc xspec}, absorbed by both simple and partial covering absorption. To reduce
the effects of cross-calibration uncertainties the \mekal\ normalisations were
permitted to optimise independently for the three EPIC instruments. All other
parameters were kept the same for the three detectors. Using two \mekal s of
different temperatures gave a \chisq\ of 1530 (\rchisq=1.20); adding a third
\mekal\ reduced \chisq\ to 1428 (\rchisq=1.12), and a fourth gave a \chisq\ of
1331 (\rchisq=1.05). The addition of a fifth \mekal\ did not improve the fit.

The {\sc cemekl} model in {\sc xspec} reproduces \mekal\ emission along a
continuous range of temperatures, where normalisation varies according to
$(T/T_{\rm MAX})^\alpha$. We tried replacing our four discrete \mekal s with a
{\sc cemekl}, but this worsened the fit to a \chisq\ of 1385 (\rchisq=1.08).

\begin{figure*}
\begin{center}
\psfig{file=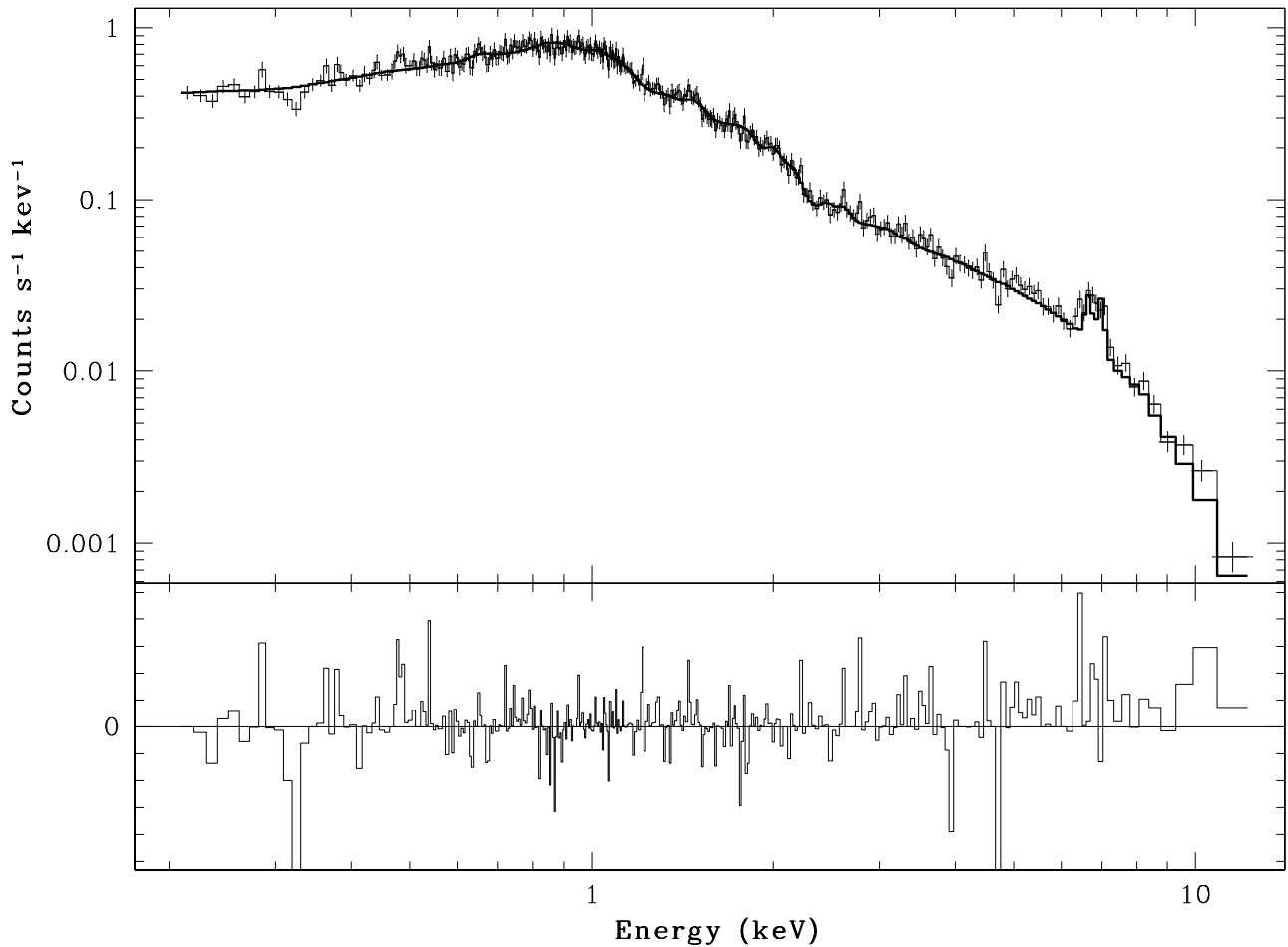,width=17cm}
\caption{The EPIC-pn spectrum of HT~Cam, fitted by a four-\mekal\ model (see
text). Note the prominence of the thermal Iron-K emission complex around 6.7
keV. The residuals provide evidence for a 6.4-keV component of cold iron,
having an equivalent width of 70 eV.}
\label{fig:spec}
\end{center}
\end{figure*}

In all of the above fits, the column density of the partial absorber tended to
zero. We thus removed it, with no affect on the fit quality. We also tried
adding a soft blackbody component, as is seen in some IPs (e.g.\ de~Martino
\etal2004) but this gave almost no improvement in fit quality
($\Delta\chisq\til4$). The EPIC-pn spectrum plus the four-\mekal\ model are
shown in Fig.~\ref{fig:spec}, and the parameters for the model are given in 
Table~\ref{tab:spec}. Note that, as the parameters are expected to vary on the
spin cycle, these may be weighted averages.

\begin{figure}
\begin{center}
\psfig{file=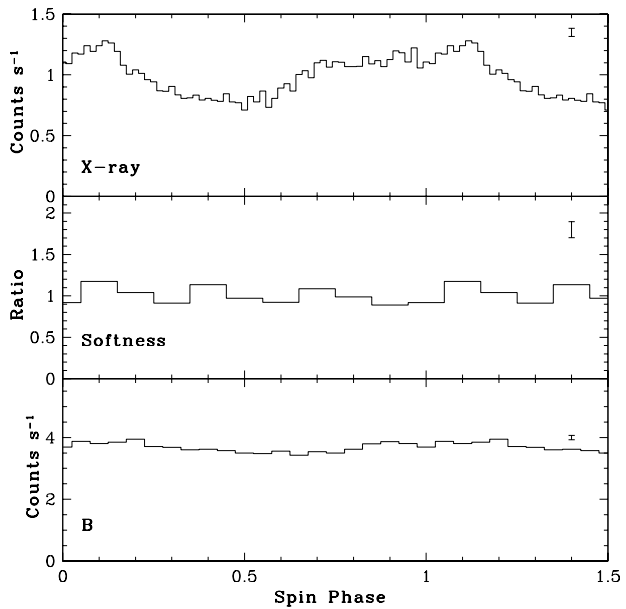,width=8.1cm}
\caption{The spin pulse-profile of HT~Cam, with typical errors. The upper
panel shows the mean 0.2--12 keV flux from all three EPIC cameras. The softness
ratio is defined as (0.2--4)/(6--12) keV. The $B$ band is as Fig.~1. Phase zero
is defined from the ephemeris of Kemp \etal(2002).}
\label{fig:spin}
\end{center}
\end{figure}

\begin{table}
\begin{center}
\begin{tabular}{lccc}
\hline
Component &  Parameter                & Value               & Error \\ 
          &  (Units) \\
\hline
Absn.     &  $n_{\rm H}$ (\cms)       & 5.35\tim{20}        & (+0.28, $-$0.45) \\ 
Mekal     &  $kT$ (keV)               & 0.172               & (+0.029, $-$0.033) \\
          &  Abundance                & 0.622               & (+0.103, $-$0.104) \\
          &  Norm (M1)                & 4.11\tim{-5}        & (+8.05, $-$4.11) \\
          &  Norm (M2)                & 7.63\tim{-5}        & (+9.94, $-$4.67) \\
          &  Norm (PN)                & 1.02\tim{-4}        & (+6.24, $-$3.32) \\
Mekal     &  $kT$ (keV)               & 0.678               & (+0.022, $-$0.022) \\
          &  Norm (M1)                & 1.52\tim{-4}        & (+0.33, $-$2.67) \\
          &  Norm (M2)                & 1.46\tim{-4}        & (+0.32, $-$2.54) \\
          &  Norm (PN)                & 1.97\tim{-4}        & (+6.53, $-$0.14) \\
Mekal     &  $kT$ (keV)               & 1.99                & (+0.29, $-$0.26) \\
          &  Norm (M1)                & 5.45\tim{-4}        & (+1.69, $-$2.09) \\
          &  Norm (M2)                & 6.88\tim{-4}        & (+0.93, $-$1.72) \\
          &  Norm (PN)                & 5.59\tim{-4}        & (+1.38, $-$1.33) \\
Mekal     &  $kT$ (keV)               & 13.4                & (+2.6, $-$2.1) \\
          &  Norm (M1)                & 2.25\tim{-3}        & (+0.67, $-$0.16) \\
          &  Norm (M2)                & 2.00\tim{-3}        & (+0.17, $-$0.10) \\
          &  Norm (PN)                & 2.06\tim{-3}        & (+8.23, $-$0.12) \\
\hline
\end{tabular}
\caption{Model components and parameters fitted to the phase-averaged spectrum
of HT~Cam (see Section~\ref{sec:spec}). The errors are given to the same power
of ten as the values.}
\label{tab:spec}
\end{center}
\end{table}


\section{The spin pulse}
\label{sec:spin}

The spin-pulse profile of HT~Cam is shown in Fig.~\ref{fig:spin}. We can
describe the X-ray pulse as a flat maximum covering phases \til0.7--1.1 and a
dip covering \til0.2--0.6. To investigate the cause of this dip we extracted
spectra from the two phase regions and fitted the four-\mekal\ model to both
simultaneously. Initially we allowed all parameters to vary between the two
regions, which gave a \chisq\ of 1750 (\rchisq=1.04). This was almost unchanged
(\chisq=1753, \rchisq=1.04) if the absorption was tied between the regions. We
tried adding a partial absorber to the phase-minimum spectrum, but it made no
difference to the fit, with the column density tending to zero. As a check we
tried forcing the emission to remain constant across spin-phase, varying only
the absorption. This gave a poor fit (\chisq=1969 \rchisq=1.15). We thus
conclude that the spin-pulse is not caused by absorption variations, but must
result from variations in the underlying X-ray emission. 

The softness ratio shows almost no modulation on the spin-period, implying that
the modulation is energy-independent. We thus fixed the absorption and the
ratios of the emission components across spin phase, attempting to reproduce
the observed spectral changes by reducing the flux during the dip but keeping
the `colour' constant. This gave a good fit (\chisq=1780, \rchisq=1.05) which
suggests that the modulation is indeed energy independent within \xmm's
pass-band.

In an IP the accretion disc is magnetically truncated, with material flowing
along the magnetic field lines in large `accretion curtains'. In many systems
these curtains periodically pass through our line of sight, causing a prominent
absorption dip (e.g.\ AO~Psc, Hellier, Cropper \&\ Mason 1991; FO~Aqr,
Beardmore \etal1998; V1223~Sgr, Taylor \etal1997). The low absorption in our
spectroscopy and the energy-independent nature of the spin-pulse mean that this
cannot be happening in HT~Cam. Further, they suggest that the post-shock
accretion columns themselves have an unusually low opacity, so that we see no
difference in the spectra when viewing them at different angles. This contrasts
with, for example, V405~Aur, where the accretion curtains never cross our line
of sight, but the changing aspect of the columns themselves gives
hardness-ratio changes on the spin-period (Evans \&\ Hellier 2004). 

We suggest that, in contrast, the spin pulse in HT~Cam arises solely from
simple occultation effects as the accretion columns pass over the white-dwarf
limb as it rotates. This allows us to investigate the X-ray spin pulse
with a simple geometrical model.

\section{Modelling the effects of occultation}

We constructed a model which consists of arc-shaped accretion regions of zero
height above the white-dwarf surface (since the majority of the X-ray emission
in an IP occurs close to the white-dwarf surface, see Cropper, Wu \&\ Ramsay
2000). The shape and location of the arcs was determined by assuming a disc
disrupted at $r_{\rm mag}$, and by tracing field lines from there to the
white-dwarf surface, assuming a magnetic dipole inclined to the spin axis by an
angle $\delta$ and offset from the white-dwarf centre in an arbitrary
direction.  A schematic diagram of this model is shown in Fig.~\ref{fig:schem}.
Since accretion rate will likely vary with azimuth, being greatest at the point
to which the magnetic pole points, we assume a brightness varying as

\begin{equation}
B(\gamma)=A\cos^2 \left(\frac{\pi\gamma}{2\Delta\gamma}\right)
\end{equation}

\noindent where $\gamma$ is the magnetic longitude of a point on the footprint,
between end-points $\pm\Delta\gamma$, and $A$ is a brightness coefficient,
which can take different values for the upper and lower regions. By default the
upper accretion lay between $\pm90\deg$ and the lower one covered 90--270\deg.
The star was then rotated through one revolution in 100 phase bins and a
simulated pulse-profile was produced by summing the brightness of the visible
emitting regions. We ignored projection effects, treating the X-ray emission as
optically thin.

Some form of asymmetry between the two accretion regions is necessary to
produce any modulation, otherwise the occultation of the upper pole is
compensated for by the emergence of the lower pole. We explored three sources
of asymmetry: offsets of the dipole in different directions; reducing the
azimuthal extent of one region compared to the other; and reducing the
brightness of one region compared to the other. We explored these asymmetries
over a full range of system inclination ($i$) and the dipole inclination
($\delta$), in steps of 10\deg.

A selection of model pulse-profiles is given in Fig.~\ref{fig:mod} and
discussed below. Note that if we reverse the sense of the asymmetry between the
poles, the pulse-profiles are inverted. 

\subsection{Offsetting the dipole}

\begin{figure}
\begin{center}
\psfig{file=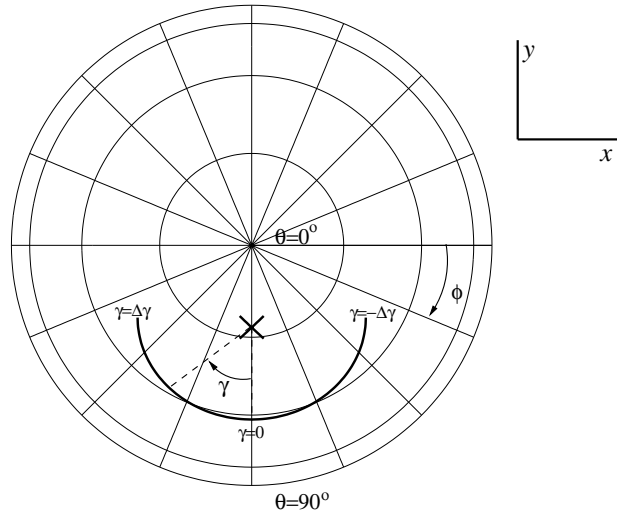,width=8.1cm}
\caption{A schematic diagram of the model used, shown here at phase zero for a
system inclination of $i=0$\deg\ and a dipole inclination of $\delta=20$\deg.
The upper magnetic pole is marked by a cross. The bold arc shows the accretion
footprint covering $-$90\deg\ to +90\deg\ of magnetic longitude. The $x$ and
$y$ axes are marked. The $z$ axis lies perpendicular to the page, with positive
$z$ coming `towards' the reader. Geometric latitude ($\theta$) and longitude
($\phi$), and magnetic longitude ($\gamma$) are shown.}
\label{fig:schem}
\end{center}
\end{figure}

\begin{figure*}
\begin{center}
\psfig{file=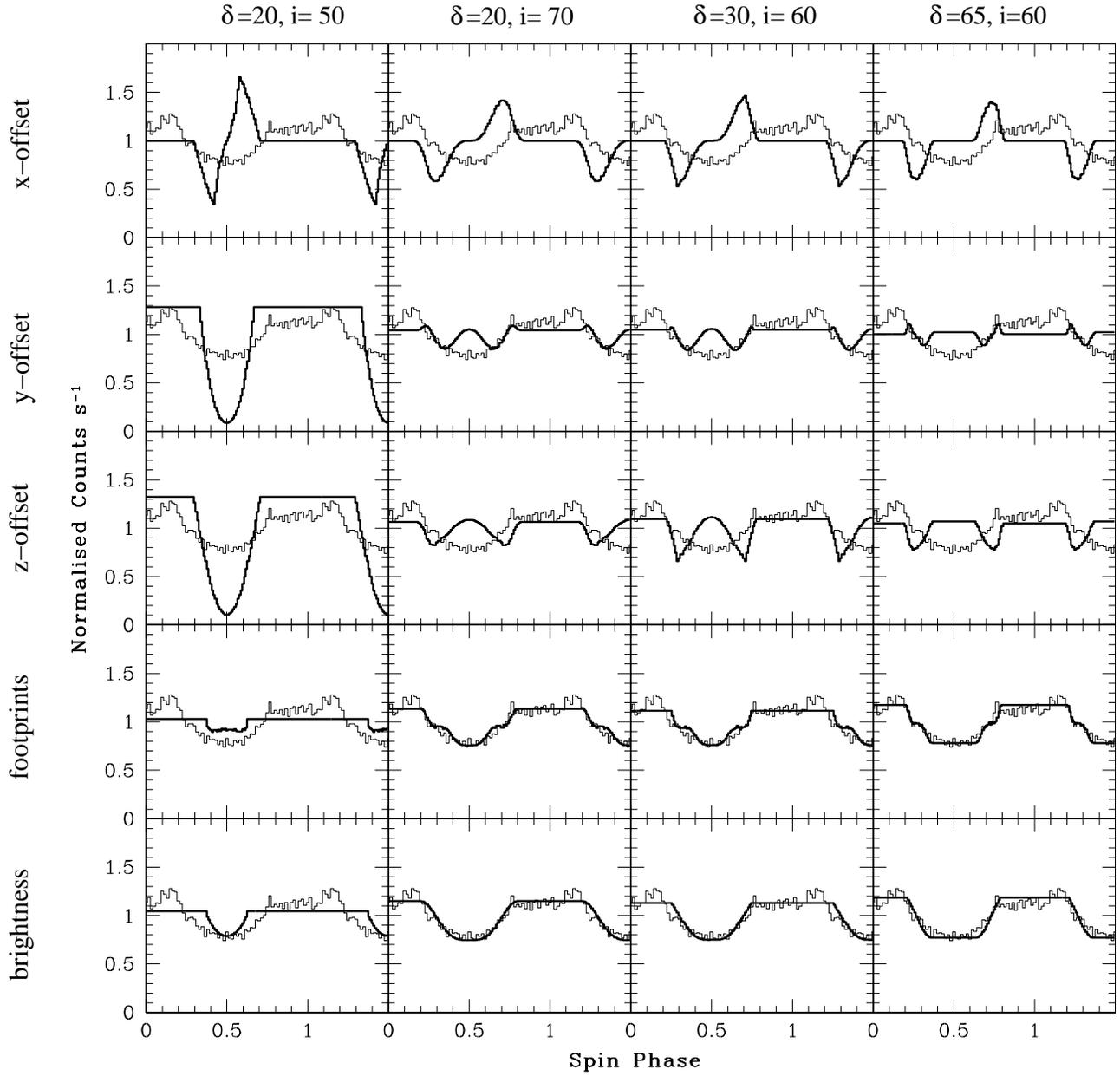,width=17cm}
\caption{Simulated X-ray pulse profiles. In the upper rows the dipole is
offset by 0.15 \rwd\ in the $x$, $y$- and $z$ directions. The upper
footprint covers $-90\deg\leq\gamma\leq90\deg$ and the lower ones covers
$90\deg\leq\gamma\leq270\deg$ in all but the `footprints' row. Here
the lower footprint covers $120\deg\leq\gamma\leq240\deg$. In the bottom row
the lower footprint is 65 per cent as bright as the upper one.}
\label{fig:mod}
\end{center}
\end{figure*}

We first offset the dipole along the $x$-axis (perpendicular to the plane of
the spin and magnetic axes, see Fig.~\ref{fig:schem}). Since disappearance of
one pole is not coincident with appearance of the other, there is a modulation.
However it is very different from the data (Fig.~\ref{fig:mod}, top row), so is
not considered further.

If the offset is instead in the $y$-direction (in the plane of the spin and
magnetic axes), the lower footprint will not appear over the limb at all when
$i$ and \d\ are low. This results in a deep dip (Fig.~\ref{fig:mod}, second
row). The flux seen at minimum in Fig.~\ref{fig:mod} results because the upper
pole is never totally occulted.  When $i$ and \d\ are higher the lower
footprint does appear, but not until after the upper one has begun to be
occulted, hence a partial dip is seen. Once the lower pole does appear it
causes the flux to rise, causing a `hump' during the dip.

Invoking a $z$-offset (i.e.\ along the spin axis) makes the magnetic latitude
of the footprints bigger at one pole than the other. At low values of $i$ and
\d\ this prevents the lower footprint from ever being seen (Fig.~\ref{fig:mod},
third row). As $i$ and \d\ increase, the lower footprint does appear, but as it
is smaller than the upper one it appears and disappears more rapidly, which
gives rise to a `hump'.

None of the above offsets result in a pulse profile resembling that in HT~Cam.

\subsection{Reducing the size of the lower pole}

We tried reducing the longitudinal range of the lower arc, keeping the
brightness coefficient ($A$) the same, so that the lower pole becomes fainter
overall. The difference in total brightness between the poles gives rise to a
dip. The mid-part of ingress is flat, since appearance of one pole is
cancelling with disappearance of the other. However, the effect of the smaller
footprint is confined more to the middle of the ingress, so a decline is seen at
the beginning and end of ingress.

In the above the total brightness of the lower footprint is less than that of
the upper. We have explored, instead, changing the brightness coefficient such
that the total brightnesses are the same. This gives a worse match to the data,
with a `hump' being seen instead of a dip, and is not shown in
Fig.~\ref{fig:mod}.

Again, the asymmetries considered in this section do not produce profiles like
that seen in HT~Cam.

\subsection{Making the lower footprint fainter}

Keeping the dipole symmetrical, but reducing the brightness coefficient for
the lower footprint, results in a smooth modulation as the upper footprint is
progressively replaced with the fainter lower one (Fig.~\ref{fig:mod}, bottom
row). Changing the system and dipole inclinations alters the width and profile
of the dip. If \d\ and $i$ are high the entire lower pole is visible for
some time, producing a flat-bottomed minimum. 

This model gives the only good match to the data. Good fits can be found
for a system inclination of \til40--70\deg, and a dipole inclination of
\til20--50\deg. Here the footprints lie between $\pm90$\deg\ (upper) and
90--270\deg\ (lower). To reproduce the depth of the dip the lower footprint is
65 per cent as bright as the upper one. The magnetic latitude of these regions
corresponds to a magnetospheric radius of \til1.3--3.6 \rwd.
Fig.~\ref{fig:best} shows the system with the best-fitting parameters
($i=55\deg$, $\d=35\deg$, $R_{\rm mag}=2.6\rwd$) and shows that the accretion
curtains do not intersect our line of sight in this geometry, as required by
the lack of absorption in the spectra.


\section{Discussion}
\label{sec:disc}

\begin{figure*}
\begin{center}
\hbox{
\psfig{file=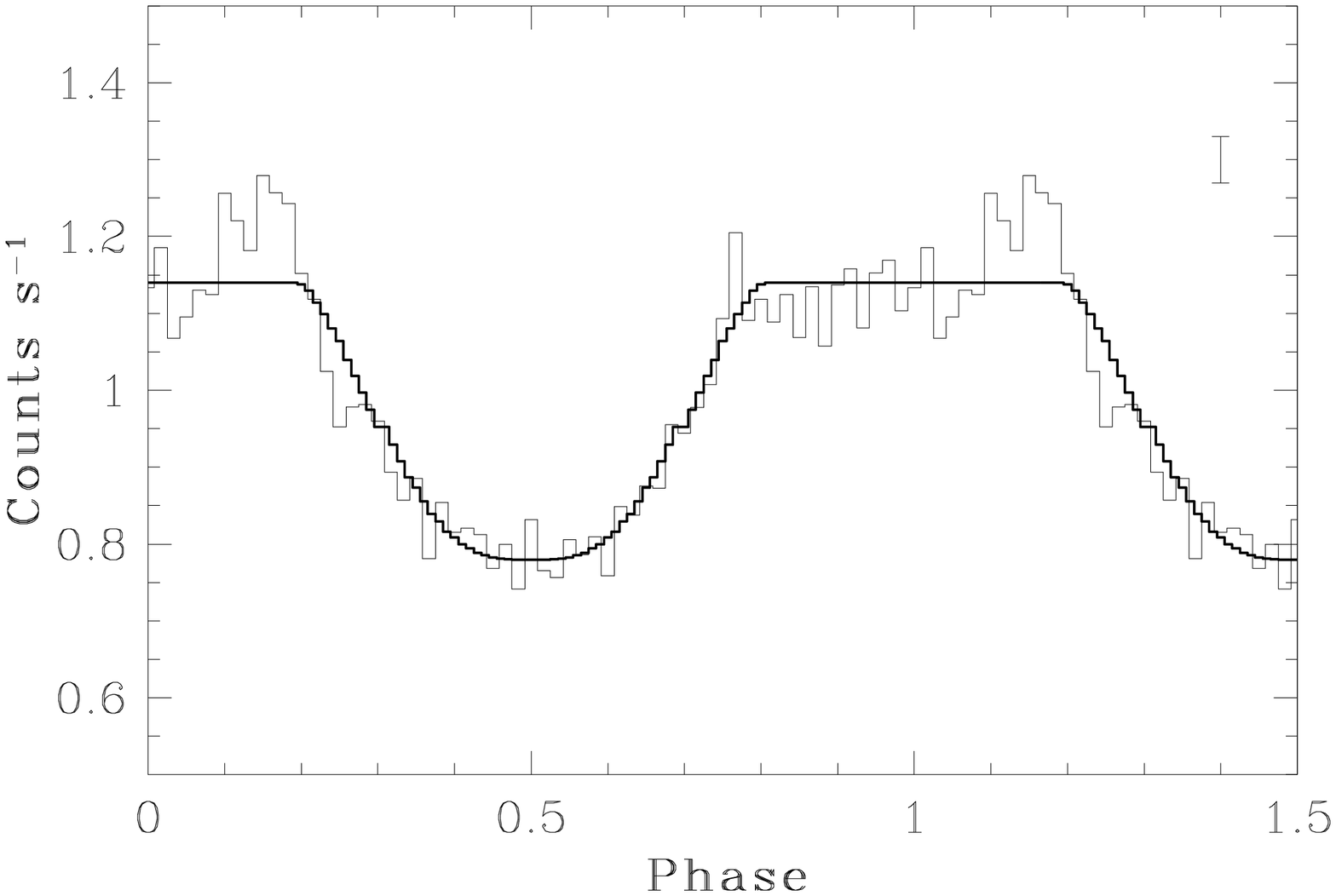,width=8.1cm}
\hspace{1.5cm}
\psfig{file=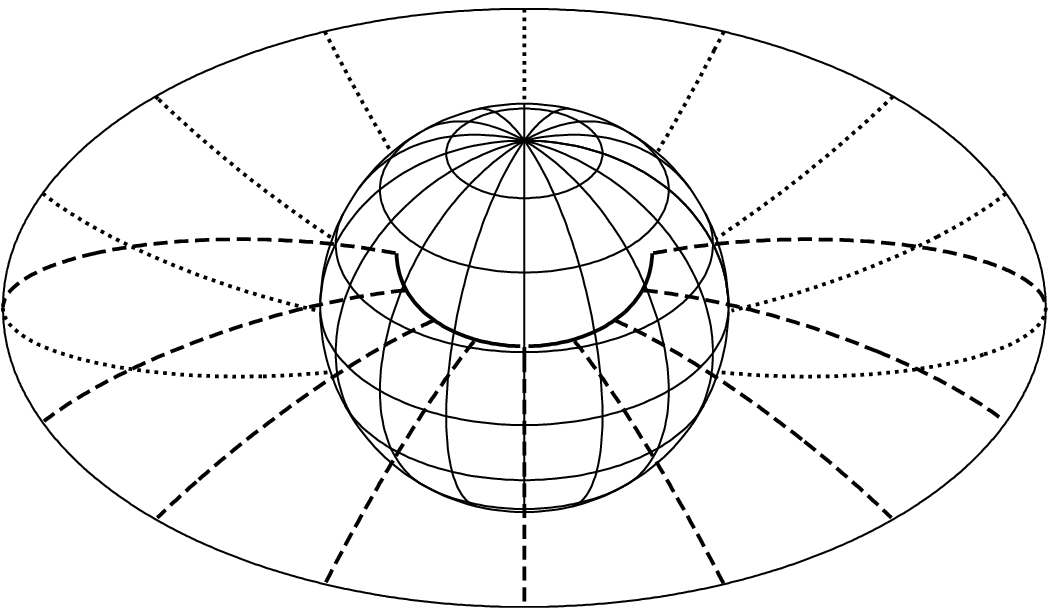,width=8.1cm}
}
\caption{The best-fitting model, as described in the text. \emph{Left panel:}
The data and model lightcurve. \emph{Right panel:} A schematic view of the
system. The upper and lower accretion curtains are shown as dashed and dotted
lines respectively, while the ring shows the inner edge of the disc. We can see
that the accretion curtains never cross our line of sight to the footprints.}
\label{fig:best}
\end{center}
\end{figure*}

We have found there to be little evidence for opacity in the X-ray spectrum of
HT~Cam, suggesting that the spin pulse must result solely from geometric
effects. We have thus produced model pulse profiles for a range of possible
geometries. We find that only one matches the data, namely one in which the
lower accretion region is fainter than the upper one. By fine-tuning this model
to the data we find that the inclination of HT~Cam is \til55\deg, while the
magnetic  dipole makes an angle of \til35\deg\ to the spin axis (see
Fig.~\ref{fig:best}).

Analysis of an \xmmn\/\ observation of HT~Cam has shown the pulse-profile to be
energy-independent. By creating model pulse-profiles we show that this can
be explained by the upper pole being periodically occulted by the white-dwarf,
provided that the lower accretion region is fainter than the upper one.  Our
models also suggest that the inclination of HT~Cam is \til55\deg, while the
magnetic  dipole makes an angle of \til35\deg\ to the spin axis. 

The energy independent nature of the X-ray spin-pulse in HT~Cam and the lack of
absorption in its spectrum is unusual, since most IPs show spectral variations
over the spin cycle and prominent, phase-varying absorption. One part of the
explanation is that, in HT~Cam, the accretion curtains do not cross our
line-of-sight to the X-ray emitting regions. But this alone is not sufficient
to explain the energy-independent X-ray modulation: Evans \&\ Hellier (2004)
and de~Martino \etal(2004) report that the same is true in V405~Aur, however
the pulse profile of that star is energy dependent. In that system, the energy
dependence is thought to arise from opacity within the accretion columns, which
changes with the viewing angle.

The fact that we don't see this in HT~Cam suggests that the opacity in the
accretion column is low. Only two other IPs are known in which a dense absorber
($n_H > 2\tim{21}$ \cms) is not needed to fit the X-ray spectrum, namely EX~Hya
and V1025~Cen (see Allan, Hellier \&\ Beardmore 1998 and Hellier, Beardmore \&\
Buckley 1998, respectively). Like HT~Cam, these system both lie below the
period gap, where systems are believed to have accretion rates an order of
magnitude lower than those above the gap (e.g.\ Warner 1987). 

Mukai \etal(2003) came to a similar conclusion for EX~Hya. They analysed
\emph{Chandra\/} HETG spectra of several IPs above the gap and also of EX~Hya.
They found much more line emission in the latter, which they attributed to an
optically thin accretion column, probably resulting from  a low specific
accretion rate.

We thus suggest that the X-ray spectra of IPs depend on whether they are above
or below the cataclysmic variable period gap. Those above the period gap have
high accretion rates, resulting in optically thick accretion columns and
prominent phase-varying absorption when accretion curtains obscure the line of
sight. Those below the gap have much lower accretion rates, optically thin
accretion columns and show much less absorption in their spectra.

\section*{Acknowledgements}

We thank Sean Harmer for his help during this analysis.

\label{lastpage}

\begin{thebibliography}{}


\bibitem{}
Allan A., Hellier C., Beardmore A.P., 1998, MNRAS, 295, 167

\bibitem{} 
Beardmore A.P., Mukai K., Norton A.J., Osborne J.P., Hellier C.,
1998, MNRAS, 297, 337

\bibitem{}
Cropper M., Wu K., Ramsay G., 2000, NewAR, 44, 57

\bibitem{}
de~Martino D., Matt G., Belloni T., Haberl F., Mukai K., 2004, A\&A,
415, 1009

\bibitem{}
de~Martino D., Matt G., Mukai K., Bonnet-Bidaud J.M., G\"ansicke B.T., Haberl
F., Mouchet M., 2005, in Hameury J.M., Lasota J.P., eds, ASP Conf. Ser Vol.
330, The astrophysics of cataclysmic variables and related objects, Astron.
Soc. Pac., San Francisco, p. 403

\bibitem{}
Evans P.A., Hellier C., 2004, MNRAS, 353, 447

\bibitem{}
Hellier C., 1997, MNRAS, 291, 71

\bibitem{}
Hellier C., Cropper M., Mason K.O., 1991, MNRAS, 248, 233

\bibitem{}
Hellier C., Beardmore A.P., Buckley D.A.H., 1998, MNRAS, 299, 851

\bibitem{}
Jansen F. et al., 2001, A\&A, L1

\bibitem{}
Kemp J., Patterson J., Thorstensen J.R., Fried R.E., Skillman D., Billings G.,
2002, PASP, 114, 623

\bibitem{}
Mason K.O., 1997, MNRAS, 285, 493 

\bibitem{}
Mason K.O., et al., 2001, A\&A, 365, L36

\bibitem{}
Mukai K., Kinkhabwala A., Peterson J.R., Kahn S.M., Paerels F., 2003, ApJ, 586,
L77

\bibitem{}
Patterson J., 1994, PASP, 106, 209

\bibitem{}
Str\"{u}der L. et al., 2001, A\&A, 365, L18

\bibitem{}
Taylor P., Beardmore A.P., Norton A.J., Osborne J.P., Watson M.G., 1997, MNRAS,
289, 349

\bibitem{}
Tovmassian G.H., et al., 1998, A\&A, 335, 227

\bibitem{}
Turner M.J.L. et al., 2001, A\&A, 365, L27

\bibitem{}
Warner B., 1987, MNRAS, 227, 23

\end{thebibliography}
\end{document}